\documentclass[conference]{IEEEtran}
\IEEEoverridecommandlockouts
\usepackage{cite}
\usepackage{amsmath,amssymb,amsfonts}
\usepackage{algorithm}
\usepackage{algorithmic}
\usepackage{graphicx}
\usepackage{textcomp}
\usepackage{xcolor}
\usepackage{subfigure}
\def\BibTeX{{\rm B\kern-.05em{\sc i\kern-.025em b}\kern-.08em
    T\kern-.1667em\lower.7ex\hbox{E}\kern-.125emX}}
\begin{document}

\title{Energy Efficiency of Rate-Splitting Multiple Access for Multibeam Satellite Communications\\
\thanks{The research was supported by A*STAR under its RIE2020 Advanced Manufacturing and Engineering (AME) Industry Alignment Fund - Pre Positioning (IAF-PP) under Grant No. A19D6a0053. Yao Ge was supported by the RIE2020 Industry Alignment Fund - Industry Collaboration Projects (IAF-ICP) Funding Initiative, as well as contributions from the industry partner(s).}
}

%%The work of J. Liu and Y. L. Guan was supported by A*STAR under its RIE2020 Advanced Manufacturing and Engineering (AME) Industry Alignment Fund - Pre Positioning (IAF-PP) under Grant No. A19D6a0053.

\author{

        % <-this % stops a space
% <-this % stops a space% <-this % stops a space
}

\author{\IEEEauthorblockN{Jinyuan Liu\IEEEauthorrefmark{1}, Yong Liang Guan\IEEEauthorrefmark{1}, Yao Ge\IEEEauthorrefmark{2}, Longfei Yin\IEEEauthorrefmark{3}, and Bruno Clerckx\IEEEauthorrefmark{3}\IEEEauthorrefmark{4} 
}
\IEEEauthorblockA{\IEEEauthorrefmark{1}School of Electrical and Electronic Engineering,
Nanyang Technological University, Singapore}
\IEEEauthorblockA{\IEEEauthorrefmark{2}Continental-NTU Corporate Lab, Nanyang Technological University, Singapore}
\IEEEauthorblockA{\IEEEauthorrefmark{3}Department of Electrical and Electronic Engineering, Imperial College London, London, UK}
\IEEEauthorblockA{\IEEEauthorrefmark{4}Silicon Austria Labs (SAL), Graz A-8010, Austria}
Emails: jinyuan001@e.ntu.edu.sg, \{eylguan;yao.ge\}@ntu.edu.sg, \{longfei.yin17;b.clerckx\}@imperial.ac.uk
}

\maketitle

\begin{abstract}
Energy efficiency (EE) problem has become an important and major issue in satellite communications. In this paper, we study the beamforming design strategy to maximize the EE of rate-splitting multiple access (RSMA) for the multibeam satellite communications by considering imperfect channel state information at the transmitter (CSIT). We propose an expectation-based robust beamforming algorithm against the imperfect CSIT scenario. By combining the successive convex approximation (SCA) with the penalty function transformation, the nonconvex EE maximization problem can be solved in an iterative manner. The simulation results demonstrate the effectiveness and superiority of RSMA over traditional space-division multiple access (SDMA). Moreover, our proposed beamforming algorithm can achieve better EE performance than the conventional beamforming algorithm.
\end{abstract}


\section{Introduction}
Owing to the explosive increase of wireless communication devices in 5G and beyond, satellite communication has played an important role due to its ubiquitous coverage and ability to provide service in unserved areas \cite{b1}. 
To increase the spectral efficiency (SE) and handle the inter-group interference problem of multibeam satellite communications, there are some research works applying space-division multiple access (SDMA) based beamforming design strategy to maximize the sum rate of the whole system \cite{b2,b3}. However, SDMA cannot support massive user connectivity and is sensitive to the quality of channel state information at the transmitter (CSIT). Hence, to further improve the SE and handle the interference problem efficiently, rate-splitting multiple access (RSMA) based beamforming design strategy for multigroup multicast scenario is first proposed in \cite{b4}, where RSMA achieves better max-min fairness (MMF) rate performance than SDMA. Motivated by \cite{b4}, the authors in \cite{b5} further extend the MMF rate maximization study of RSMA to the multibeam satellite communications with different CSIT qualities.

However, the aforementioned works only consider SE maximization while neglecting the energy efficiency (EE) aspect, which is a crucial issue under the background of economic and operational considerations of satellites, as well as environmentally friendly transmission behaviors. Recently, the EE maximization problem in the satellite scenario has been studied in \cite{b6,b7} based on the SDMA strategy. Additionally, motivated by the advantage of RSMA in SE, there are also plenty of works applying RSMA for the EE problem under terrestrial communication system \cite{b8,b9,b10}. However, all these works mentioned above assume perfect CSIT, which is not practical due to the imperfect and outdated CSIs feedback from the users to the satellite or to the base station. 
% In\cite{b5}, Dinkelbach’s method is used to handle the fractional programming based on zero forcing (ZF). A two-step quadratic transformation is proposed in \cite{b6} to transform the EE maximization problem into an equivalent convex one and solved by alternating optimization algorithm. 
Hence, the robust EE design of RSMA for multibeam satellite communications is desired and promising to support future ubiquitous coverage of next-generation access networks. 

In this paper, in contrast to \cite{b8,b9,b10} that focus on perfect CSIT, we aim to maximize the EE by considering a more practical and general imperfect CSIT setting of multibeam satellite communications. To handle the original highly nonconvex problem, we apply successive convex approximation (SCA) to combine with a penalty function transformation. As a result, the nonconvex EE maximization problem can be efficiently transferred into a convex one and solved in an iterative manner. To the best of our knowledge, this is the first work to study the EE maximization problem of RSMA for the multibeam satellite communication scenario with imperfect CSIT. Our EE simulation results demonstrate the robustness and superiority of RSMA over SDMA. Furthermore, our proposed beamforming design algorithm can achieve better EE performance than conventional Dinkelbach's weighted minimum mean square error (D-WMMSE) algorithm \cite{b11}. 
\section{System Model}
% We consider a Ka-band multibeam geostationary (GEO) satellite communication system, where a satellite equipped with $N_t$ antennas communicates with $K$ single-antenna users. The users are grouped into $M (1 \leq M \leq K)$ multicast groups, and the $\mathcal{G}_{m}$ denotes the set of users belonging to group $m$, for all $m \in \mathcal{M}=\{1 \cdots M\}$. We assume each user $k$ is served by only one beam, thus, $\mathcal{G}_{i} \cap \mathcal{G}_{j}=\emptyset$, for all $i,j \in \mathcal{M}, i\ne j$. We construct the widely applied single feed per beam (SFPB) architecture \cite{b4}, which means one beam can be generated by one feed (i.e. $N_t=M$). We also assume feeder link to be an ideal channel and adopt full frequency reuse to improve the spectral efficiency. The detailed system design including transmitter, channel and receiver models will be given in this section. 
We consider a Ka-band multibeam geostationary (GEO) satellite communication system, where a satellite equipped with $N_t$ antennas communicates with $K$ single-antenna users. The set of all user indices is defined as $\mathcal{K}=\{1 \cdots K\}$. The users are grouped into $M (1 \leq M \leq K)$ multicast groups, and we denote $\mathcal{G}_{m}$ as the set of users belonging to group $m$, for all $m \in \mathcal{M}=\{1 \cdots M\}$. With the assumption that each user is served by only one beam, we construct the widely applied single feed per beam (SFPB) architecture \cite{b4}, in which one beam can be generated by one feed (i.e. $N_t=M$). We also assume the feeder link as an ideal channel and adopt full frequency reuse to improve SE. The detailed system design including transmitter, channel, and receiver models will be given in this section. 
\subsection{Transmitter Model}
In this paper, we apply the RSMA transmission scheme to the multibeam satellite system. At the transmitter, each message is divided into common part and private part. Then, all common part messages are grouped together and encoded into one common stream denoted by $s_c$, and the remaining private part messages are encoded into private stream independently, where $s_m$ denotes the $m$-th group private stream. Hence, there are overall $M+1$ symbol streams, which is $\mathbf{s}=[s_c,s_1,...,s_M]^T \in \mathbb{C}^{(M+1) \times 1}$, and we assume $\mathbb{E}\left\{\mathbf{s s}^{H}=\mathbf{I}\right\}$. Then, all the streams are linearly precoded by the beamforming matrix $\mathbf{W}=\left [\mathbf{w}_c, \mathbf{w}_1,...\mathbf{w}_M   \right ] \in \mathbb{C}^{N_t \times (M+1)}$, which gives the transmitted signal as
\begin{equation}
\mathbf{x}=\mathbf{w}_{c} s_{c}+\sum_{m=1}^{M} \mathbf{w}_{m} s_{m},
\label{eq:transsignal}
\end{equation}
%Then all the streams are linearly precoded by the precoding matrix $\mathbf{W}=\left [\mathbf{w_c}, \mathbf{w_1},...\mathbf{w_M}   \right ] \in \mathbb{C}^{N_t \times (M+1)}$, which gives the transmitted signal as
where $\mathbf{w}_{c} \in \mathbb{C}^{N_t \times 1}$ and $\mathbf{w}_{m} \in \mathbb{C}^{N_t \times 1}$ denote the corresponding common and $m$-th group's beamformer, respectively. The maximum transmission power is set as $P_t$, thus, the total transmit power constraint can be given by $\operatorname{tr}\left(\mathbf{W} \mathbf{W}^{H}\right) \leq P_{t}$. 

For the total power consumption at the satellite, we use a practical model that the power consumption contains two parts including rate-dependent and rate-independent power consumption. Hence, the total power consumption can be written as
\begin{equation}
\overline{P}(\mathbf{W})=\frac{1}{\beta }  \operatorname{tr}\left(\mathbf{W} \mathbf{W}^{H}\right)+\mathrm{P}_{\mathrm{c}}+\xi \overline{R}(\mathbf{W}),
\label{eq:powerconsmodel}
\end{equation}
where $\beta \in [0,1]$ denotes the power amplifier efficiency, $\mathrm{P}_{\mathrm{c}}$ is the rate-independent circuit power consumption, and $\xi \ge 0$ represents the power consumption per unit data rate. The ergodic data rate $\overline{R}(\mathbf{W})$ will be introduced and defined in Section \ref{c} later on.

\subsection{Channel Model}
After being processed at the satellite, the signal will propagate through the channel. Here, we model the channel between the satellite and $k$-th user as \cite{b5}
\begin{equation}
\mathbf{h}_{k}= \mathbf{r }_{k} \odot \exp \left\{j \boldsymbol\theta_{k}\right\} \odot \mathbf{b}_{k} 
\label{eq:channel},
\end{equation}
where $\odot$ denotes the Hadamard product. $\mathbf{r}_{k}\in \mathbb{C}^{N_t \times 1}$ is the rain attenuation whose element can be defined as $r_{k, n_{t}}=\psi_{k, n_{t}}^{1 / 2}$, and its dB form $\psi_{k, n_{t}}^{dB}=20{\log_{10}{\psi_{k, n_{t}}}}$, commonly follows log-normal random distribution, i.e., $\ln \left(\psi_{k,n_{t}}^{d B}\right) \sim \mathcal{N}(\mu, \sigma^2)$. $\boldsymbol\theta_{k} = \left [ \theta_{k,1}, \theta_{k,2},..., \theta_{k,N_t} \right ]^{T}\in \mathbb{C}^{N_t \times 1} $ is the channel phase vector with each element uniformly distributed between $0$ and $2 \pi$. Moreover, $\mathbf{b}_{k}\in \mathbb{C}^{N_t \times 1}$ is the coefficient composed of the large-scale fading and satellite beam radiation pattern, whose $n_t$-th element is
\begin{equation}
b_{k, n_{t}}= \frac{v \sqrt{G_{R} G_{k, n_{t}}}}{4\pi f d_{k} \sqrt{\kappa T_{s y s} B_{w}}},
\label{eq:b}
\end{equation}
with $f$ being carrier frequency, $v$ being the light speed, and $d_{k}$ giving the distance between the $k$-th user and satellite. $G_{R}$ is the receive antenna gain of the user, $\kappa$ is the Boltzmann constant; $T_{s y s}$ and $B_{w}$ denote the receive noise temperature and carrier bandwidth, respectively. Additionally, $G_{k, n_{t}}$ represents the beam radiation pattern which can be modeled as 
\begin{equation}
G_{k, n_{t}}=G_{\max }\left[\frac{J_{1}\left(u_{k, n_{t}}\right)}{2 u_{k, n_{t}}}+36 \frac{J_{3}\left(u_{k, n_{t}}\right)}{u_{k, n_{t}}^{3}}\right]^{2}.
\label{eq:G}
\end{equation}
In (\ref{eq:G}), $G_{\max }$ is the maximum satellite beam gain, $J_{1}(.)$ and $J_{3}(.)$ denote the first and third order of first-kind Bessel functions, respectively. Furthermore, $u_{k, n_{t}}=2.07123 \times \left [ \sin \left(\phi_{k, n_{t}}\right) / \sin \left(\phi_{3 \text{dB}}\right)  \right ]$ with $\phi_{k, n_{t}}$ being the angle between $k$-th user and the $n_t$-th beam center, $\phi_{3 \text{dB}}$ being the 3-dB loss angle compared with beam center.

It can be observed that the amplitude of the satellite channel is determined by the large-scale fading factor, which can be considered as a constant and easily detected during the intervals of interest. However, owing to the fog and atmospheric absorption, the channel phase is time-varying and changing very fast, which makes it difficult to obtain the perfect value \cite{b12}. Hence, we give the relationship between real phase $\boldsymbol\theta_k$ and estimated phase $\overline{\boldsymbol\theta}_k$ as
\begin{equation}
\boldsymbol\theta_k={\overline{\boldsymbol\theta}_k}+\mathbf{e}_k,
\label{eq:phase}
\end{equation}
with $\mathbf{e}_k=\left [e_{k,1}, e_{k,2},..., e_{k,N_t} \right ]^{T}$ being the phase error vector following the distribution $\mathbf{e}_{k} \sim \mathcal{N}\left(\mathbf{0}, \delta^{2} \mathbf{I}\right)$, where $\delta^{2}$ denotes phase error variance. Since we assume the amplitude of the channel is perfectly known, then the channel vector between the satellite and $k$-th user can be expressed as
\begin{equation}
\mathbf{h}_{k}=\overline{\mathbf{h}}_{k} \odot \mathbf{q}_{k}=\operatorname{diag}\left(\overline{\mathbf{h}}_{k}\right) \mathbf{q}_{k},
\label{eq:imperfectchannel}
\end{equation}
where $\overline{\mathbf{h}}_{k}$ denotes the estimated channel vector, and $\mathbf{q}_{k} \triangleq \exp \left\{j \mathbf{e}_{k}\right\}$.

% where $\odot$ denotes the Hadamard product. $\mathbf{r}_{k}\in \mathbb{C}^{N_t \times 1}$ is the rain attenuation whose element can be defined as $r_{k, n_{t}}=\psi_{k, n_{t}}^{1 / 2}$, and its dB form $\psi_{k, n_{t}}^{dB}=20{\log_{10}{\psi_{k, n_{t}}}} $ commonly follows log-normal random distribution, i.e., $\ln \left(\psi_{k,n_{t}}^{d B}\right) \sim \mathcal{N}(\mu, \sigma)$. $\boldsymbol\theta_{k} = \left [ \theta_{k,1}, \theta_{k,2},..., \theta_{k,N_t} \right ]^{T}\in \mathbb{C}^{N_t \times 1} $ is the channel phase vector with each element uniformly distributed between $0$ and $2 \pi$. Moreover, $\mathbf{b}_{k}\in \mathbb{C}^{N_t \times 1}$ is the coefficient composed of the large-scale fading and satellite beam radiation pattern, whose $n_t$-th element is
\subsection{Receiver Model}\label{c}
% In this paper, RSMA transmission scheme is applied to the multibeam satellite system. At the transmitter, there are total $M$ messages $F_1,...,F_M$ for the $M$ groups of users respectively. Based on RS scheme, each message is divided into common part and private part, i.e. $F_{m} \rightarrow\left\{F_{m, c}, F_{m, p}\right\}$. Then all common part messages are grouped together and encoded into a common stream shared by all groups, i.e. $\left\{F_{1, c}, ... F_{M, p}\right\}  \rightarrow s_c$, and the remaining private part messages are encoded into private stream independently, i.e. $F_{m, p} \rightarrow s_m$. Hence, there are overall $M+1$ symbol streams are transmitted, which is $\mathbf{s}=[s_c,s_1,...,s_M]^T \in \mathbb{C}^{(M+1) \times 1}$, where we assume $\mathbb{E}\left\{\mathbf{s s}^{H}=\mathbf{I}\right\}$. 
After propagating through the channel, the received signal at the $k$-th user can be expressed as
\begin{equation}
y_{k}=\mathbf{h}_{k}^{H} \mathbf{w}_{c} s_{c}+\mathbf{h}_{k}^{H} \mathbf{w}_{\mu(k)} s_{\mu(k)}+\mathbf{h}_{k}^{H} \sum_{m=1, m \neq \mu(k)}^{M} \mathbf{w}_{m} s_{m}+n_{k},
\label{eq:receivignal}
\end{equation}
where $\mu(k) \in \mathcal{M}$ is the mapping function between the $k$-th user and its group index. $n_{k} \sim \mathcal{C N}\left(0, \sigma_{n}^{2}\right)$ denotes the Additive White Gaussian Noise (AWGN) at $k$-th user, and we assume equal noise variances $\sigma_{n}^{2}$ for all users. Then, each user firstly decodes the common stream and regards the remaining private streams as noise. The SINR of decoding common stream at $k$-th user is
\begin{equation}
\Gamma_{c, k}=\frac{\left|\mathbf{h}_{k}^{H} \mathbf{w}_{c}\right|^{2}}{\left|\mathbf{h}_{k}^{H} \mathbf{w}_{\mu(k)}\right|^{2}+\sum_{m=1, m \neq \mu(k)}^{M}\left|\mathbf{h}_{k}^{H} \mathbf{w}_{m}\right|^{2}+\sigma_{n}^{2}}.
\label{eq:commonsir}
\end{equation}
In our work, to handle the imperfect CSIT, we assume only statistical knowledge of the channel is available at the satellite. Consequently, the corresponding ergodic common rate can be defined as $ \overline{R}_{c,k} \triangleq \mathbb{E}\left\{\log _{2}(1+\Gamma_{c, k})\right\}$, and to ensure each user can decode the common stream, the shared common rate is determined by the weakest user as following
\begin{equation}
\overline{R}_{c} \triangleq \min _{k \in \mathcal{K}} \overline{R}_{c, k}.
\label{eq:commonrate}
\end{equation}
Since the common stream is shared among all the groups, then $\overline{R}_{c} \triangleq  {\textstyle \sum_{m=1}^{M} C_m}$, where $C_m$ denotes the portion of common rate allocated to group $m$. Having successfully decoded and removed the common stream via successive interference cancellation (SIC), each user decodes its private stream by treating the inter-group interference as noise. The SINR of decoding private stream for $k$-th user is determined by
\begin{equation}
\Gamma_{p,k}=\frac{\left|\mathbf{h}_{k}^{H} \mathbf{w}_{\mu(k)}\right|^{2}}{\sum_{m=1, m \neq \mu(k)}^{M}\left|\mathbf{h}_{k}^{H} \mathbf{w}_{m}\right|^{2}+\sigma_{n}^{2}}. 
\label{eq:privatesnr}
\end{equation}
The corresponding ergodic private rate can be written as $\overline{R}_{p,k} \triangleq \mathbb{E}\left\{\log _{2}(1+\Gamma_{p,k})\right\}$. To ensure each user in group $m$ can decode the multicast information, the $m$-th private group rate $\overline{R}_m$ is determined by the weakest user in the same group and given by
\begin{equation}
\overline{R}_m \triangleq \min _{k \in \mathcal{G}_{m}} \overline{R}_{p,k}. 
\label{eq:privaterate}
\end{equation}
Hence, the overall ergodic transmission rate for the satellite system can be written as
\begin{equation}
\overline{R}(\mathbf{W})= {\textstyle \sum_{m=1}^{M} C_m+\overline{R}_m}. 
\label{eq:allrate}
\end{equation}

%%----------------------------------------------------
\section{Problem Formulation and Proposed Solution}

\subsection{Problem Formulation}
In this section, we formulate the EE maximization problem of RSMA for multibeam satellite communications as
\begin{subequations}\label{e}
\begin{align}
\mathop {\max }\limits_{\left\{\mathbf{W}, \mathbf{c}, \mathbf{r}_m \right\}}\quad &\mathop  \mathrm{E}(\mathbf{W})\label{e1}\\
\text{s.t.}\quad &{\textstyle \sum_{m=1}^{M}} C_{m} \leq \overline{R}_{c, k}, \quad \forall k \in \mathcal{K}  \label{e2}\\
&C_{m} \geq 0, \quad \forall m \in \mathcal{M} \label{e3} \\
&\overline{R}_{p, k } \ge  \overline{R}_{m},  \forall k \in \mathcal{G}_{m} \label{e4} \\
&C_{m}+\overline{R}_{m} \geq R_{\mathrm{th}}, \quad \forall m \in \mathcal{M} \label{e5} \\
&\operatorname{tr}(\mathbf{W}\mathbf{W}^{\mathrm{H}}) \le P_{t} \label{e6},
\end{align}
\end{subequations}
where $\mathbf{c}=[C_1,...,C_M]$ and $\mathbf{r}_{m}=[\overline{R}_1,...,\overline{R}_M]$. The objective function is the global EE \cite{b13} defined as 
\begin{equation}
\mathrm{E}(\mathbf{W})= \frac{\overline{R}(\mathbf{W})}{\overline{P}(\mathbf{W})}. 
\label{eq:GEE}
\end{equation}
The constraints (\ref{e2}) and (\ref{e4}) are to ensure each user can decode the common stream and multicast stream, respectively. Constraint (\ref{e3}) ensures allocated common rate for each group is nonnegative. (\ref{e5}) and (\ref{e6}) are the QoS constraint and total transmission power constraint, respectively. Note that the problem (\ref{e}) is a nonconvex fractional program with nonconvexity mainly from rate expressions, which makes the problem difficult to solve. 

% Based on the ergodic transmission rate model and total transmission model built in the previous section, we define the global EE \cite{b10} of the whole system shown as
% \begin{equation}
% \mathrm{E}(\mathbf{W})= \frac{\overline{R}(\mathbf{W})}{\overline{P}(\mathbf{W})},
% \label{eq:GEE}
% \end{equation}
% where the global EE is defined as the ergodic achievable rate divided by the average consumed energy. Thus, the EE maximization problem considering imperfect CSIT can be formulated as

\subsection{Proposed Solution}
To address the nonconvex fractional program, we first define $\mathbf{W}_c=\mathbf{w}_c \mathbf{w}_c^H$, $\mathbf{W}_m=\mathbf{w}_m \mathbf{w}_m^H$, and $\mathbf{W}_s=[\mathbf{W}_c,\mathbf{W}_1,...,\mathbf{W}_M]$. We also introduce slack variables $\boldsymbol\alpha=\left \{x,y \right \}$ and $z$ to equivalently transform the original problem (\ref{e}) into a semi-definite programming (SDP) form with rank-one constraints, which is shown as follows
\begin{subequations}\label{16}
\begin{align}
% {\mathop{\max}\limits_{\left\{\mathbf{W_c},\mathbf{W_m}, \right. \\ \left. \mathbf{c}, \mathbf{r}_{m},\boldsymbol\alpha \right\}}}\ &\mathop z \label{e21}\\
{\mathop{\max}\limits_{\left\{\mathbf{W}_s, \mathbf{c} \hfill\atop 
\mathbf{r}_{m},\boldsymbol\alpha \right\}}}\ &\mathop z \label{e21}\\
\text { s.t. } \ &\frac{x^2}{y} \ge z \label{e22}\\
&\overline{R}(\mathbf{W}_s) \ge x^2 \label{e23}\\
&y \ge \overline{P}(\mathbf{W}_s) \label{e24}\\
%& \overline{R}_{p,k} \ge r_k, \forall k \in \mathcal{K} \label{e25}\\
%& r_k \ge R_m, \forall k \in \mathcal{G}_m, \forall m \in \mathcal{M} \label{e25.5}\\
&\operatorname{tr}\left(\mathbf{W}_{c}\right)+   {\textstyle \sum_{m=1}^{M}}  \operatorname{tr}\left(\mathbf{W}_{m}\right) \leq P_{t} \label{e26}\\
&\mathbf{W}_{c} \succeq 0, \mathbf{W}_{m} \succeq 0, \quad \forall m \in \mathcal{M} \label{e27}\\
&\operatorname{rank}\left(\mathbf{W}_{c}\right)=\operatorname{rank}\left(\mathbf{W}_{m}\right)=1, \forall m \in \mathcal{M} \label{e28}\\
&(\text{\ref{e2}})-(\text{\ref{e5}}) \label{e29}.
\end{align}
\end{subequations}
% \begin{subequations}
% \begin{align}&\mathop {\max }\limits_{{\bf{G}},{\bf{v}},{\bf{u}},{t_D},{t_R},{t_S}\hfill\atop
% \scriptstyle\lambda, \bf{E},\mathcal{D},\mathcal{B},{\bf{\chi }}}\quad \mathop F\left( {\lambda ,{\bf{G}},{\bf{v}},{t_S},{t_D},{t_R}} \right)\\
% &\text{s.t.}\quad 2\ln {E_D} - {\beta _D} + 1 \ge \frac{{{t_D}}}{{{\alpha _D}}},\\
%  &2\ln {E_R} - {\beta _R} + 1 \ge \frac{{{t_R}}}{{{\alpha _R}}},\\
%   &2\ln {E_M} - {\beta _M} + 2\ln {E_t} - {\beta _t} + 2\ln {E_d} - {\beta _d} + 3 \ge 0,\\
%   &(\ref{Robust_S_R}),(\ref{Robust_S_power}),(\ref{Robust_S_u}),(\ref{PD_SR}),(\ref{PD_M}),(\ref{PD_POWER}),(\ref{PD_susis_D}),(\ref{PD_ALTER}),(\ref{PD_susis_Dd})
% \end{align}
% \end{subequations}
Note that the constraint (\ref{e22}) is still in a fractional form which is nonconvex. Hence, we apply the first-order Taylor series approximation at the left side of (\ref{e22}), which is given by 
\begin{equation}
\frac{x^{2}}{y} \geq \frac{2 x^{[n]}}{y^{[n]}} x-\left(\frac{x^{[n]}}{y^{[n]}}\right)^{2} y  \ge z\triangleq \Omega^{[n]}(x, y) \ge z. 
\label{eq:sca}
\end{equation}
% In this case the constraint (\ref{e22}) can be rewritten as
% \begin{equation}
% \Omega^{[n]}(x, y) \ge z. 
% \label{eq:sca}
% \end{equation}

Then, our remaining challenge is to tackle the nonconvexity of constraints (\ref{e27})-(\ref{e29}). We first deal with the nonconvex rate constraints (\ref{e2}) and (\ref{e4}). Since the rate expressions are in their ergodic form, we can approximate them by taking the expectation operation into the log expression. Hence, by taking the common rate as an example, $\overline{R}_{c, k}$ can be approximated shown as follow 
%Since the rate expressions are in its ergdoic form, we can approximate them by referring the following transformation: for random variables $X$ and $Y$, the following approximation holds
%\begin{equation}
%\mathbb{E}\left\{\log _{2}\left(1+\frac{\mathrm{X}}{\mathrm{Y}}\right)\right\} \approx \log _{2}\left(1+\frac{\mathbb{E}\{\mathrm{X}\}}{\mathbb{E}\{\mathrm{Y}\}}\right),
%\label{eq:approximation}
%\end{equation}
%where this approximation has been verified to be theoretically accurate in \cite{b11}. Hence, by taking the common rate as an example, $\overline{R}_{c, k}$ can be approximated shown as follow 
\begin{equation}
\begin{aligned}
& \overline{R}_{c,k}=\mathbb{E}\left\{\log _{2}\left(1+\Gamma_{c, k}\right)\right\}\\
&\approx \log _{2}\left(\frac{\mathbb{E}\left\{\operatorname{tr}\left(\mathbf{H}_{k} \mathbf{W}_{c}\right)\right\}+\sum_{m=1}^{M} \mathbb{E}\left\{\operatorname{tr}\left(\mathbf{H}_{k} \mathbf{W}_{m}\right)\right\}+\sigma_{n}^{2}}{\sum_{m=1}^{M} \mathbb{E}\left\{\operatorname{tr}\left(\mathbf{H}_{k} \mathbf{W}_{m}\right)\right\}+\sigma_{n}^{2}}\right)\\
&=\log _{2}\left(\frac{\operatorname{tr}\left(\overline{\mathbf{H}}_{k} \mathbf{W}_{c}\right)+\sum_{m=1}^{M} \operatorname{tr}\left(\overline{\mathbf{H}}_{k} \mathbf{W}_{m}\right)+\sigma_{n}^{2}}{\sum_{m=1}^{M} \operatorname{tr}\left(\overline{\mathbf{H}}_{k} \mathbf{W}_{m}\right)+\sigma_{n}^{2}}\right).
\end{aligned}
\label{appr}
\end{equation}
This approximation has been verified to be theoretically accurate in \cite{b14}. Based on the (\ref{eq:imperfectchannel}), $\overline{\mathbf{H}}_{k}$ in (\ref{appr}) is defined as
\begin{equation}
\overline{\mathbf{H}}_{k} =\operatorname{diag}\left(\overline{\mathbf{h}}_{k}\right) \overline{\mathbf{Q}}_{k} \operatorname{diag}\left(\overline{\mathbf{h}}_{k}^{\mathrm{H}}\right),
\end{equation}
where $\overline{\mathbf{Q}}_{k}= \mathbb{E} \left \{\mathbf{q}_{k} \mathbf{q}_{k}^{\mathrm{H}} \right \}$. Then, based on the approximated rate expression, we introduce slack variables $\eta_{p,k}, \chi_{p,k}, \eta_{c, k}, \chi_{c, k}$ and expand the rate constraints (\ref{e2}) and (\ref{e4}) as follows
\begin{subequations}
\begin{align}
&\eta_{c, k}-\chi_{c, k} \geq  {\textstyle \sum_{m=1}^{M} }  C_{m} \log 2, \forall k \in \mathcal{K} \label{D21c}\\
&e^{{\eta}_{c, k}} \leq \operatorname{tr}\left(\overline{\mathbf{H}}_{k} \mathbf{W}_{c}\right)+ \overline{I}_{c,k}+\sigma_{n}^{2}, \forall k \in \mathcal{K} \label{D21d}\\
&e^{\chi_{c,k}} \geq \overline{I}_{c,k}+\sigma_{n}^{2}, \forall k \in \mathcal{K} \label{D21e}\\
&\eta_{p,k}-\chi_{p,k} \geq \overline{R}_{m} \log 2, \forall k\in \mathcal{G}_{m}, \forall m\in \mathcal{M} \label{D21f}\\
&e^{{\eta}_{p,k}} \leq \operatorname{tr}\left(\overline{\mathbf{H}}_{k} \mathbf{W}_{\mu\left(k\right)}\right)+\overline{I}_{p,k}+\sigma_{n}^{2}, \forall k \in \mathcal{K} \label{D21g}\\
&e^{\chi_{p,k}} \geq \overline{I}_{p,k} +\sigma_{n}^{2}, \forall k \in \mathcal{K} \label{D21h}
\end{align}
\end{subequations}
%Then, based on the approximated rate expression and (\ref{eq:sca}), we introduce slack variables $\eta_{p,k}, \chi_{p,k}, \eta_{c, k}, \chi_{c, k}$ and expand the rate constraints (\ref{e2}) and (\ref{e4}), the problem (\ref{16}) can be approximately transformed into the following problem 
% \begin{subequations}
% \begin{align}
% & \max _{\left\{\mathbf{W_c},\mathbf{W_m}, \mathbf{c},\mathbf{r}_m, \boldsymbol\alpha, \boldsymbol\lambda \right\}} \ z \label{D21}\\
% \text { s.t } \ & \Omega^{[n]}(x, y) \ge z \label{D21b}\\
% &\eta_{c, k}-\chi_{c, k} \geq  {\textstyle \sum_{m=1}^{M} }  C_{m} \log 2, \forall k \in \mathcal{K} \label{D21c}\\
% &e^{{\eta}_{c, k}} \leq \operatorname{tr}\left(\overline{\mathbf{H}}_{k} \mathbf{W}_{c}\right)+ \overline{I}_{c,k}+\sigma_{n}^{2}, \forall k \in \mathcal{K} \label{D21d}\\
% &e^{\chi_{c,k}} \geq \overline{I}_{c,k}+\sigma_{n}^{2}, \forall k \in \mathcal{K} \label{D21e}\\
% &\eta_{p,k}-\chi_{p,k} \geq R_{m} \log 2, \forall k\in \mathcal{G}_{m}, \forall m\in \mathcal{M} \label{D21f}\\
% &e^{{\eta}_{p,k}} \leq \operatorname{tr}\left(\overline{\mathbf{H}}_{k} \mathbf{W}_{\mu\left(k\right)}\right)+\overline{I}_{p,k}+\sigma_{n}^{2}, \forall k \in \mathcal{K} \label{D21g}\\
% &e^{\chi_{p,k}} \geq \overline{I}_{p,k} +\sigma_{n}^{2}, \forall k \in \mathcal{K} \label{D21h}\\
% & (\ref{e3}), (\ref{e5}), (\ref{e23})-(\ref{e28})
% \end{align}
% \end{subequations}
where $\overline{I}_{c,k}={\textstyle \sum_{m=1}^{M} } \operatorname{tr}\left(\overline{\mathbf{H}}_{k} \mathbf{W}_{m}\right)$ and $\overline{I}_{p,k}={\textstyle \sum_{m=1, m \neq \mu\left(k\right)}^{M}}  \operatorname{tr}\left(\overline{\mathbf{H}}_{k} \mathbf{W}_{m}\right)$ for notation simplicity. Then, we can observe that the constraints (\ref{D21e}) and (\ref{D21h}) are still nonconvex, thus, the first order Taylor series expansion is applied around $\chi_{c,k}^{(n)}$ and $\chi_{p,k}^{(n)}$, which can be given by
\begin{equation}
\overline{I}_{c,k}+\sigma_{n}^{ 2} \leq e^{\chi_{c, k}^{[n]}}\left(\chi_{c, k}-\chi_{c, k}^{[n]}+1\right),
\label{22}
\end{equation}
\begin{equation}
\overline{I}_{p,k}+\sigma_{n}^{2} \leq e^{\chi_{p,k}^{[n]}}\left(\chi_{p,k}-\chi_{p,k}^{[n]}+1\right).
\label{23}
\end{equation}
Furthermore, although the constraints (\ref{D21d}) and (\ref{D21g}) are convex, the computational complexity to handle these constraints is high since their left-hand sides are nonlinear. Hence, to obtain more efficient implementation, we approximate the constraints (\ref{D21d}) and (\ref{D21g}) to a set of second-order cone (SOC) constraints, which can be written as
\begin{equation}
\begin{aligned}
 & t_{c, k}\leq \operatorname{tr}\left(\overline{\mathbf{H}}_{k} \mathbf{W}_{c}\right)+\bar{I}_{c, k}+\sigma_{n}^{2}, \\
 & \left\|t_{c, k}+\eta_{c, k}-T_{c, k}^{[n]}, 2 \sqrt{t_{c, k}^{[n]}}\right\|_{2} \leq t_{c, k}-\eta_{c, k}+T_{c, k}^{[n]},
\end{aligned}
\label{24}
\end{equation}
\begin{equation}
\begin{aligned}
& t_{p, k} \leq \operatorname{tr}\left(\overline{\mathbf{H}}_{k} \mathbf{W}_{\mu\left(k\right)}\right)+\bar{I}_{p, k}+\sigma_{n}^{2}, \\
& \left\|t_{p, k}+\eta_{p, k}-T_{p, k}^{[n]}, 2 \sqrt{t_{p, k}^{[n]}}\right\|_{2} \leq t_{p, k}-\eta_{p, k}+T_{p, k}^{[n]},
\end{aligned}
\label{25}
\end{equation}
where $t_{c,k}$ and $t_{p,k}$ are introduced variables, and $T_{c,k}^{[n]}=\left(\log \left(t_{c,k}^{[n]}\right)+1\right)$, $T_{p,k}^{[n]}=\left(\log \left(t_{p,k}^{[n]}\right)+1\right)$ for notation simplicity.

The rest of the nonconvex challenge is owing to the rank-one constraint (\ref{e28}) on $\mathbf{W}_c$ and $\mathbf{W}_m$. Hence, we apply an iterative penalty function (IPF) into the objective function to ensure a rank-one solution can be obtained. Firstly, since the rank-one constraint implies only one eigenvalue is nonzero, the constraints (\ref{e28}) can be replaced by the following
\begin{equation}
\begin{array}{l}
\operatorname{tr}\left(\mathbf{W}_{c}\right)-\lambda_{\max }\left(\mathbf{W}_{c}\right)=0, \\
\operatorname{tr}\left(\mathbf{W}_{m}\right)-\lambda_{\max }\left(\mathbf{W}_{m}\right)=0, \forall m \in \mathcal{M}
\end{array}
\label{28}
\end{equation}
where the $\lambda_{\max }\left(\mathbf{W}_{c}\right)$ and $\lambda_{\max }\left(\mathbf{W}_{m}\right)$ denote the maximum eigenvalue of $\mathbf{W}_{c}$ and $\mathbf{W}_{m}$, respectively. Then, the penalty function can be written as follows
\begin{equation}
\begin{array}{l}
F=\rho ( \left[\operatorname{tr}\left(\mathbf{W}_{c}\right)-\lambda_{\max }\left(\mathbf{W}_{c}\right)\right] \\
+\sum_{m=1}^{M}\left[\operatorname{tr}\left(\mathbf{W}_{m}\right)-\lambda_{\max }\left(\mathbf{W}_{m}\right)\right] ),
% F=\rho\left(\left[\operatorname{tr}\left(\mathbf{W}_{c}\right)-\lambda_{\max }\left(\mathbf{W}_{c}\right)\right]\right. \\
% \left.+\sum_{m=1}^{M}\left[\operatorname{tr}\left(\mathbf{W}_{m}\right)-\lambda_{\max }\left(\mathbf{W}_{m}\right)\right]\right),
\end{array}
\label{pf1}
\end{equation}
% \begin{equation}
% \begin{array}{l}
% % F=\rho ( \left[\operatorname{tr}\left(\mathbf{W}_{c}\right)-\lambda_{\max }\left(\mathbf{W}_{c}\right)\right] \\
% % +\sum_{m=1}^{M}\left[\operatorname{tr}\left(\mathbf{W}_{m}\right)-\lambda_{\max }\left(\mathbf{W}_{m}\right)\right] ),
% F=\rho \Big(\left[\operatorname{tr}\left(\mathbf{W}_{c}\right)-\lambda_{\max }\left(\mathbf{W}_{c}\right)\right]\Big. \\
% \Bigg.+\sum_{m=1}^{M}\left[\operatorname{tr}\left(\mathbf{W}_{m}\right)-\lambda_{\max }\left(\mathbf{W}_{m}\right)\right]\Big ),
% \end{array}
% \label{pf1}
% \vspace{-4.0mm}
% \end{equation}
where parameter $\rho$ denotes the penalty factor. Obviously, the objective function is nonconvex if (\ref{pf1}) is directly inserted into (\ref{e21}). Hence, we apply an iterative method to acquire a convex expression. By taking the common part as an example, the following inequality can be given
\begin{equation}
\operatorname{tr}\left(\mathbf{W}_{c}\right)-\left(\mathbf{v}_{c, \max }^{[n]}\right)^{H} \mathbf{W}_{c} \mathbf{v}_{c, \max }^{[n]} \geq \operatorname{tr}\left(\mathbf{W}_{c}\right)-\lambda_{\max }\left(\mathbf{W}_{c}\right),
\end{equation}
where $\mathbf{v}_{c, \max}$ is the normalized eigenvector corresponding to $\lambda_{\max }\left(\mathbf{W}_{c}\right)$. Similarly, we will also have $\mathbf{v}_{m, \max}$ corresponding to $\lambda_{\max }\left(\mathbf{W}_{m}\right)$. Then, the penalty function can be rewritten as  
\begin{equation}
\begin{array}{l}
\overline{F}^{[n]}=\rho \left([\operatorname{tr}\left(\mathbf{W}_{c}\right)-(\mathbf{v}_{c, \max }^{[n]})^{H} \mathbf{W}_{c} \mathbf{v}_{c, \max }^{[n]}] \right. \\
 \left.+\sum_{m=1}^{M}[\operatorname{tr}\left(\mathbf{W}_{m}\right)-(\mathbf{v}_{m, \max }^{[n]})^{H} \mathbf{W}_{m} \mathbf{v}_{m, \max }^{[n]}]\right ).
\end{array}
\end{equation}
% \begin{equation}
% \begin{aligned}
% \overline{F}^{[n]}=&\rho \left([\operatorname{tr}\left(\mathbf{W}_{c}\right)-(\mathbf{V}_{c, \max }^{[n]})^{H} \mathbf{W}_{c} \mathbf{v}_{c, \max }^{[n]}] \right. \\
%  &\left.+\sum_{m=1}^{M}[\operatorname{tr}\left(\mathbf{W}_{m}\right)-\left(\mathbf{v}_{m, \max }^{[n]}\right)^{H} \mathbf{W}_{m} \mathbf{v}_{m, \max }^{[n]}]\right ).
% \end{aligned}
% \end{equation}
Hence, we can solve the following convex approximated problem to achieve EE maximization of the original problem (\ref{e}) at the $n$-th iteration:
\begin{subequations}\label{30}
\begin{align}
\max _{ \left \{\mathbf{W}_s,  \mathbf{c}, \mathbf{r}_{m} \hfill\atop  \boldsymbol\alpha,  \mathbf{S}  \right \} } \ & z-\overline{\mathrm{F}}^{[n]}   \\
\text { s.t. } \ &(\text{\ref{e3}}), (\text{\ref{e5}}),(\text{\ref{e23}})-(\text{\ref{e27}}),(\text{\ref{eq:sca}}),\\
    &(\text{\ref{D21c}}),(\text{\ref{D21f}}),(\text{\ref{22}})-(\text{\ref{25}}),
\end{align}
\end{subequations}
where $\mathbf{S}=\left \{\eta_{p,k}, \chi_{p,k}, \eta_{c, k}, \chi_{c, k} , t_{c,k}, t_{p,k}, \mathbf{v}_{c, \max}, \mathbf{v}_{m, \max} \right \}$. We can observe problem (\ref{30}) is convex involving only SOC and linear matrix inequality (LMI), which can be efficiently solved by CVX optimization software. Our beamforming design is summarised in \textbf{Algorithm 1}. The initialized beamformer satisfying the QoS and total power constraints is generated to ensure the feasibility, and the stop condition is built by predefining the parameter $\epsilon$, i.e., $\epsilon=0.0001$. Finally, we apply eigenvalue decomposition (EVD) to $\mathbf{W}_{c}$ and $\mathbf{W}_{m}$, and choose the eigenvector corresponding to the maximum eigenvalue as the suboptimal beamformer. 
\begin{algorithm}[t]
\caption{Beamforming Design of Multibeam Satellite for Energy Efficiency Maximization}
\centering
\label{mAO}
\begin{algorithmic}[1]
\STATE {\textbf{Initialize: } Set $n=0$, $z^{[n]}=0$, and generate initial points ($\mathbf{W}_{s}^{[n]}, \boldsymbol\alpha^{[n]},\mathbf{S}^{[n]}$)} 
\REPEAT 
\STATE Solve problem (\ref{30}) to obtain $z^{[n+1]}$ and update ($\mathbf{W}_{s}^{[n+1]},\boldsymbol\alpha^{[n+1]},\mathbf{S}^{[n+1]}$);
\STATE $n=n+1$;
\UNTIL$\left|z^{[n+1]}-z^{[n]}\right| \leq \epsilon$
\end{algorithmic}
\end{algorithm}

% \begin{algorithm}[t]
% \caption{Robust Beamforming Design of Multigroup Multicast Satellite for Energy Efficiency Maximization}
% \centering
% \label{mAO}
% \begin{algorithmic}[1]
% \STATE {\textbf{Initialize: } $n\gets 0$, $z^{[n]} \gets 0$ $\mathbf{W}_{c}^{[n]}$, $\mathbf{W}_{m}^{[n]}$,  $x^{[n]}$, $y^{[n]}$, $\chi^{[n]}$,$t^{[n]}$,$\mathbf{v}^{[n]}$} 
% \REPEAT 

% \STATE  Solve problem $\mathcal{A}_1$ to obtain the solution $\widetilde{z}$,$\widetilde{\mathbf{W}}_c, \widetilde{\mathbf{W}}_m, \widetilde{x}, \widetilde{y},\widetilde{\chi},\widetilde{t}$;
% \STATE Compute $\widetilde{\mathbf{v}}$, and update $n \gets n+1$, $\mathbf{W}_{c}^{[n]} \gets \widetilde{\mathbf{W}}_c$, $\mathbf{W}_{m}^{[n]} \gets \widetilde{\mathbf{W}}_m$,  $x^{[n]} \gets \widetilde{x}$, $y^{[n]} \gets \widetilde{y}$, $\chi^{[n]} \gets \widetilde{\chi}$,$t^{[n]} \gets \widetilde{t}$,$\mathbf{v}^{[n]} \gets \widetilde{\mathbf{v}}$,$z^{[n]}-\widetilde{z}$;
% \UNTIL$\left|z^{[n]}-z^{[n-1]}\right| \leq \epsilon$
% \end{algorithmic}
% \end{algorithm}
\section{Simulation Results}
\begin{figure*}[t]
\centering\vspace{-0.5cm}
\subfigure[\hspace{-0.5cm}]{ \label{fig1}
\includegraphics[width=2.3in,height=1.75in]{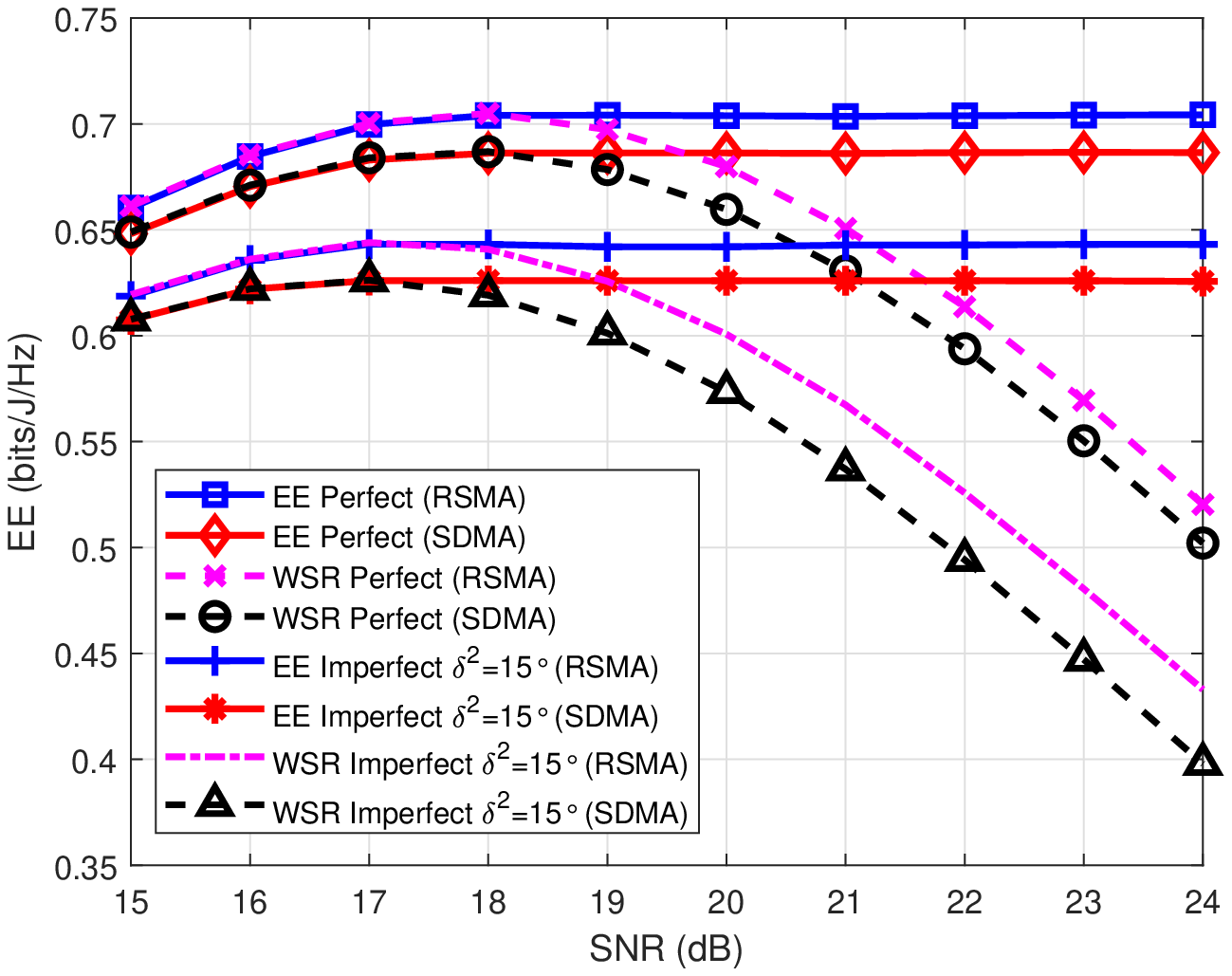}}\vspace{-1mm}
\subfigure[\hspace{-0.5cm}]{ \label{fig2}
\includegraphics[width=2.3in,height=1.75in]{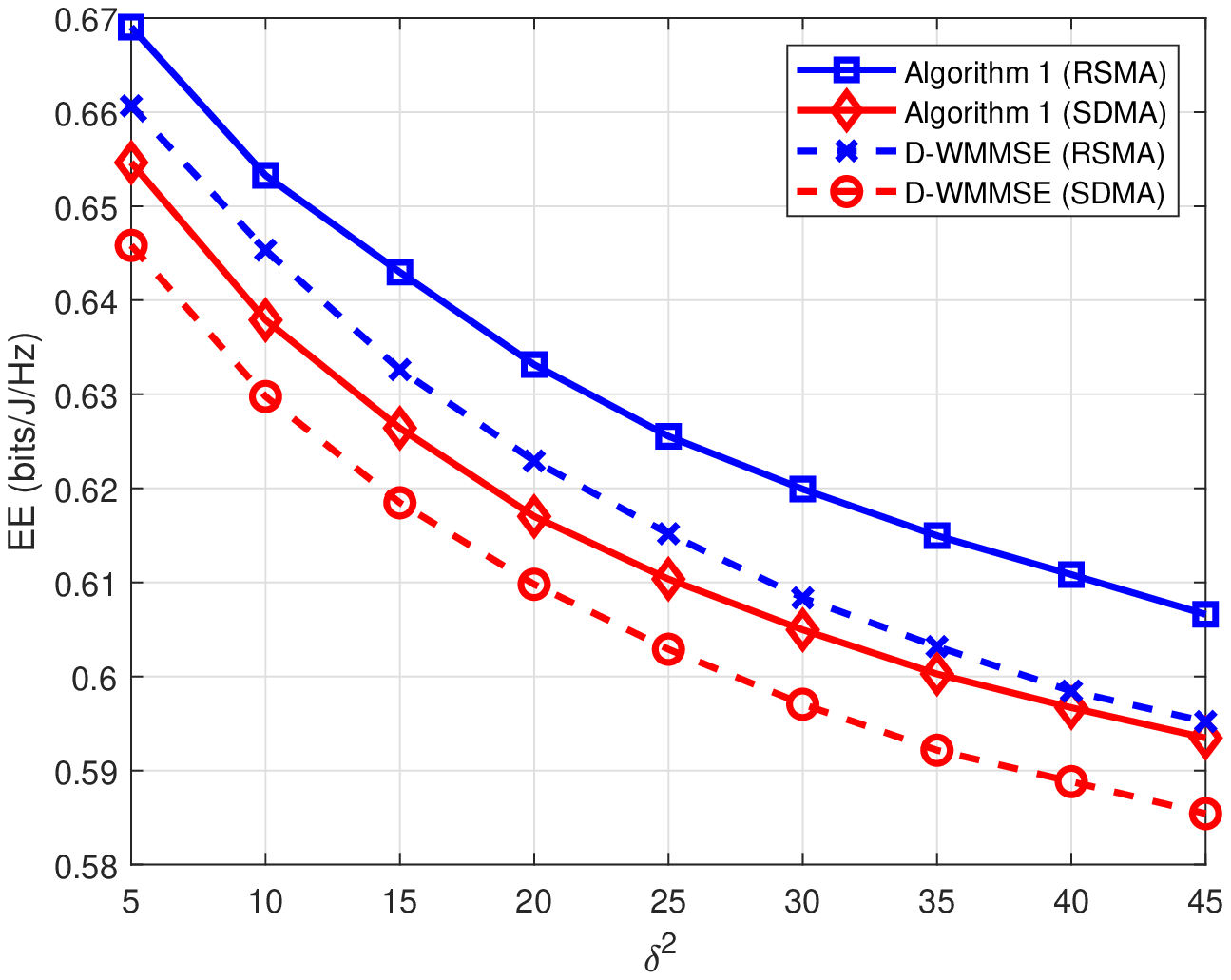}}\vspace{-1mm}
\subfigure[\hspace{-0.5cm}]{\label{fig3}
\includegraphics[width=2.3in,height=1.75in]{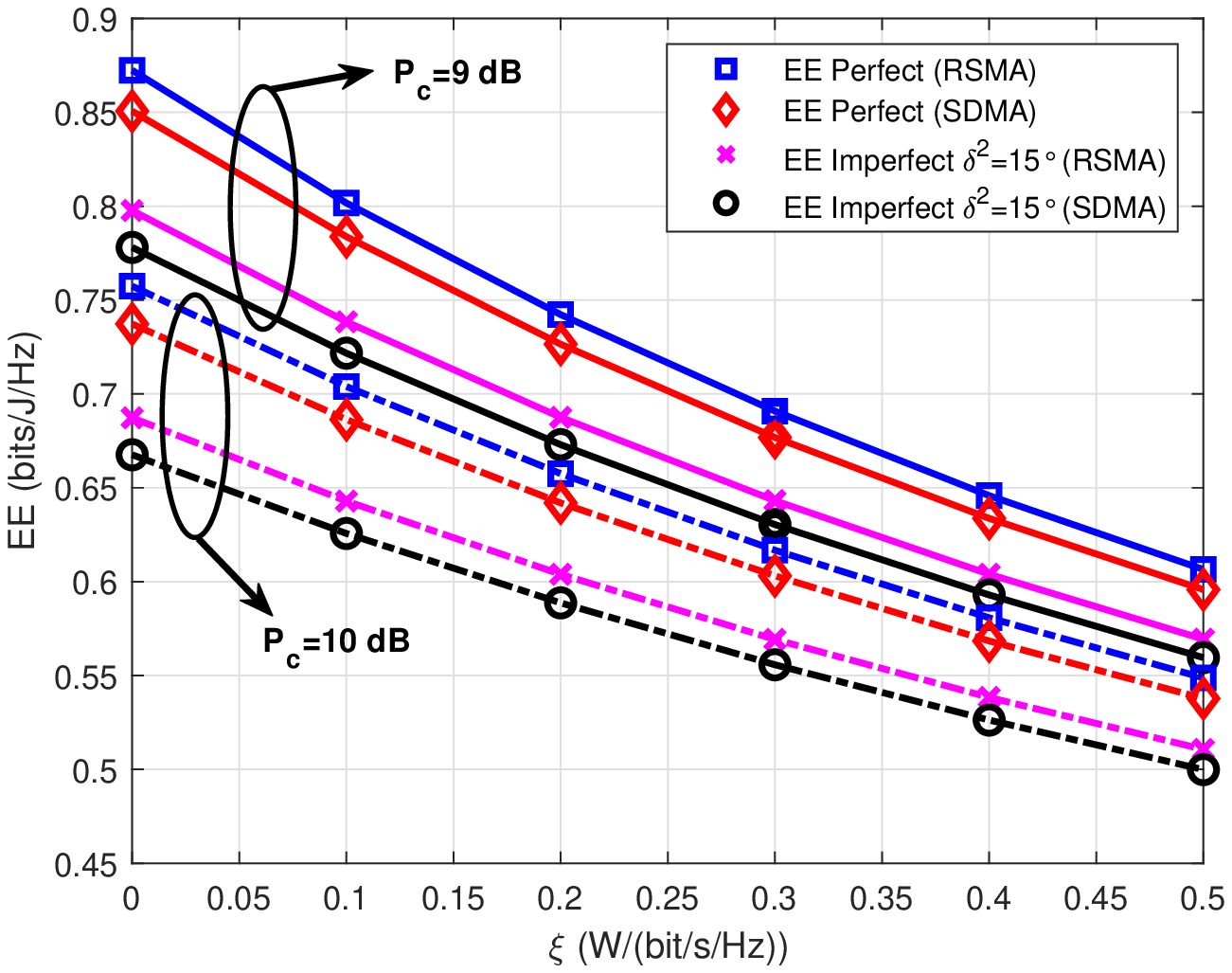}}\vspace{-1mm}
\caption{EE results of RSMA and SDMA in the GEO satellite communication system: (a) EE performance versus SNR (dB); (b) EE performance versus different phase error variances $\delta^{2}$ with different algorithms; and (c) EE performance versus $\xi$ (W/(bit/s/Hz)).}
\label{fig}
\vspace{-0.5cm}
\end{figure*}
In this section, we present the EE simulation results to evaluate our proposed RSMA beamforming algorithm. SDMA strategy and D-WMMSE algorithm are used as the benchmark. During the simulation, we assume a GEO satellite with $N_t=7$ transmitted antennas serving $M=7$ groups with two single antenna users in each group. We also define the signal-to-noise ratio (SNR) as $P_t /\sigma_n^2$, and $\beta$ is determined as $1$. The detailed system parameters are listed in Table I. All the simulation results are calculated by averaging 500 channel realizations.

\begin{table}[b]
\centering\vspace{-5.3mm}
\caption{\footnotesize System parameters for GEO SatCom system.}
\begin{tabular}{cc}
\hline
\textbf{Parameter}    & \textbf{Value}   \\ \hline
%Number of beams (groups), $N_t=M$ & $7$                  \\ 
%Number of total users, $K$ & $14$               \\ 
Carrier frequency & Ka ($20$ GHz)                            \\ 
Altitude of satellite     & $35786$ km                  \\ 
Bandwidth       & $500$ MHz               \\ 
Maximum beam gain     & $52$ dBi   \\ 
User antenna gain  & $41.7$ dBi     \\ 
Noise temperature  & $517$ K \\ 
Boltzmann's constant & $1.38 \times 10^{-23}$ J/m              \\ 
Rain fading parameters  & $(\mu, \sigma^2)=(-3.125,1.591)$                  \\ 
$3$dB angle        & $0.4^{\circ} $                  \\ 
           \hline
\end{tabular}
\label{Table1}
\end{table}

Fig.~\ref{fig1} shows the EE performance versus SNR for both perfect and imperfect CSIT of RSMA and SDMA under $ P_c=10$ dB and $\xi=0.1$ (W/(bit/s/Hz)). We can observe that RSMA outperforms SDMA in both perfect and imperfect CSIT scenarios. Moreover, we also show the EE performance when the objective function is to maximize the weighted sum rate (WSR) rather than EE, and the weights are chosen as one so that it is the same as the numerator of our EE maximization problem. It can be observed that before SNR$=18$ dB, the WSR results are alined with the EE results, but WSR results drop if SNR continually increases. This phenomenon is reasonable since our aim is to optimize the EE and keep it non-decreased when the total power increases. However, the objective of the WSR problem is to use all the available power to achieve the maximum sum rate which may sacrifice EE.  

Fig.~\ref{fig2} presents the EE performance for different beamforming algorithms with different phase error variances $\delta^{2}$ under $\text{SNR}=25$ dB, $P_c=10$ dB, and $\xi=0.1$ (W/(bit/s/Hz)). We can observe that the EE curves drop as $\delta^{2}$ increases, which is because higher phase error causes a larger channel uncertainty, and results in higher EE performance loss. However, it is clear that the achievable EE performance of our proposed \textbf{Algorithm 1} is higher than the traditional D-WMMSE algorithm, which demonstrates the robustness and advantage of our beamforming algorithm. Moreover, we can see that the higher of phase error, the bigger the EE performance gap between RSMA and SDMA, which further implies the effectiveness and robustness of RSMA compared with SDMA.  

Fig.~\ref{fig3} depicts the effect of rate-dependent and rate-independent power consumption when $\text{SNR}=25$ dB. From the results, we can notice that higher $P_c$ and $\xi$ give worse EE performance, which is reasonable since the total power consumption increases. Meanwhile, it can be observed higher value of $\xi$ gives more close results of RSMA and SDMA for all scenarios. This founding is expected since the gap between RSMA and SDMA comes from the rate improvement. When we are considering the EE problem, the increase in data rate will also cause the increase in rate-dependent power. Hence, the superiority of RSMA will decrease with the increase of $\xi$.

\section{Conclusion}
In this paper, we have studied the EE maximization problem of RSMA for multibeam satellite communications by considering imperfect CSIT. To transfer the original nonconvex problem into a convex one, we apply SCA combined with SDP and the penalty function. Hence, the original problem can be efficiently solved in an iterative manner. The EE simulation results demonstrate that RSMA outperforms SDMA under both perfect and imperfect CSIT, and our proposed beamforming algorithm is more robust compared with the traditional D-WMMSE algorithm.

% \vspace{-0.5mm}

% \bibliographystyle{ieeetr}
% \bibliography{ref}

\end{document}